\documentclass{article}
\usepackage{spconf,amsmath,graphicx}
\usepackage{amssymb}
\usepackage{hyperref}
\usepackage{booktabs}
\usepackage{threeparttable}
\usepackage{wrapfig}
\usepackage{pifont} 
\usepackage{amssymb} 
\usepackage{multirow}
\usepackage{bm}  

\title{EmoQ: Speech Emotion Recognition via Speech-Aware Q-Former and Large Language Model}
\name{Yiqing Yang$^{1}$, and Man-Wai Mak$^{1}$$^{\ast}$\thanks{$^{\ast}$ Corresponding author.}}
\address{$^{1}$Department of Electrical and Electronic Engineering,
         The Hong Kong Polytechnic University}
\begin{document}
\ninept
\setlength{\textfloatsep}{9pt}
\setlength{\floatsep}{9pt}
\maketitle

\noindent\textbf{This work has been submitted to the IEEE for possible publication. Copyright may be transferred without notice, after which this version may no longer be accessible.}
\begin{abstract}
The performance of speech emotion recognition (SER) is limited by the insufficient emotion information in unimodal systems and the feature alignment difficulties in multimodal systems. Recently, multimodal large language models (MLLMs) have made progress in SER. However, MLLMs still suffer from hallucination and misclassification problems in complex emotion reasoning. To address these problems, we propose an MLLM-based framework called EmoQ, which generates query embeddings that fuse multimodal information through an EmoQ-Former and uses multi-objective affective learning (MAL) to achieve co-optimization. The framework also provides a soft-prompt injection strategy to inject multimodal representations into the LLM. This end-to-end architecture achieves state-of-the-art performance on the IEMOCAP and MELD datasets, providing a new multimodal fusion paradigm for SER.
\end{abstract}

\section{Introduction}
\label{sec:intro}
Speech emotion recognition (SER) realizes automatic recognition of speakers' emotional states by parsing the emotional features in speech signals. However, the single modality approach has intrinsic limitations: due to the ambiguity in the definition of emotion, coupled with the complexity of linguistic expressions, it is difficult to adequately capture emotion cues at the semantic level by relying on acoustic information alone. 

Audio-based SER often lacks semantic reasoning, leading to erroneous judgments. In real-world scenarios, speech signals can be accompanied by text transcriptions through automatic speech recognition (ASR), providing a basis for multimodal fusion. Recent studies have shown that the audio-text multimodal architecture can improve the performance of emotion recognition~\cite{mmnodeformer,FMASR}. Text information can provide semantic complements to audio features, and multimodal fusion has become a key to improve SER performance.

Despite the performance gains from multimodal fusion, several critical challenges remain. First, cross-modal alignment presents a fundamental challenge~\cite{li2023blip2}. Audio signals exhibit continuous temporal dynamics, whereas transcribed texts comprise sequences of discrete tokens, creating temporal misalignment that impedes acoustic-semantic modeling. Second, modal imbalance emerges during multimodal training, where one modality tends to dominate the optimization process. Third, feature space heterogeneity poses integration difficulties~\cite{zhu2024dca}.

\begin{figure}[htbp]
\centering
\includegraphics[width=8.5cm]{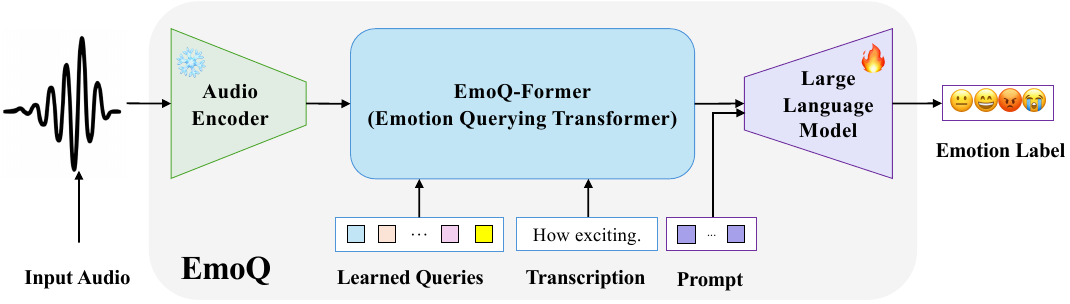}
\caption{The EmoQ framework employs an end-to-end multimodal learning approach that utilizes an LLM for emotion classification.}
\label{fig:emoq}
\end{figure}

As multimodal large language models (MLLMs) have demonstrated their capabilities in multimodal reasoning~\cite{li2023blip2,yang2024qwen2,yin2024survey}, researchers have started to apply them to SER. For example, SALMO-NN~\cite{salmonn} uses a dual-encoder architecture combined with Q-Former ~\cite{li2023blip2} and LoRA~\cite{hu2022lora} to achieve audio-text alignment. SECap~\cite{secap} further explores the possibility of using LLMs for speech emotion captioning. The authors of LLM-CL~\cite{LLM-CL} proposed a zero-shot multilingual speech emotion recognition framework based on contrastive learning. These MLLMs achieve performance gains on SER benchmarks, demonstrating the potential of the MLLM architectures for speech emotion understanding.

Inspired by the aforementioned works, we propose EmoQ as shown in Fig. \ref{fig:emoq}. EmoQ addresses the challenges in SER through three innovations:

1) EmoQ provides a cross-modal alignment mechanism, where the EmoQ-Former achieves speech-to-text feature fusion through learnable queries and attention masks to overcome temporal misalignment and feature space heterogeneity.

2) EmoQ employs a multi-objective affective learning (MAL) framework that combines supervised contrastive learning~\cite{khosla2020supervised} and focal loss~\cite{lin2017focal} for co-optimization.

3) EmoQ introduces a soft-prompt injection strategy to replace the \texttt{<AUDIO>} token in the prompt with fused representations, enabling the LLM to simultaneously utilize both language comprehension and cross-modal emotion cues for inference.

\section{Method}
\label{sec:pagestyle}
\begin{figure*}[t]
\centering
\includegraphics[width=17.8cm]{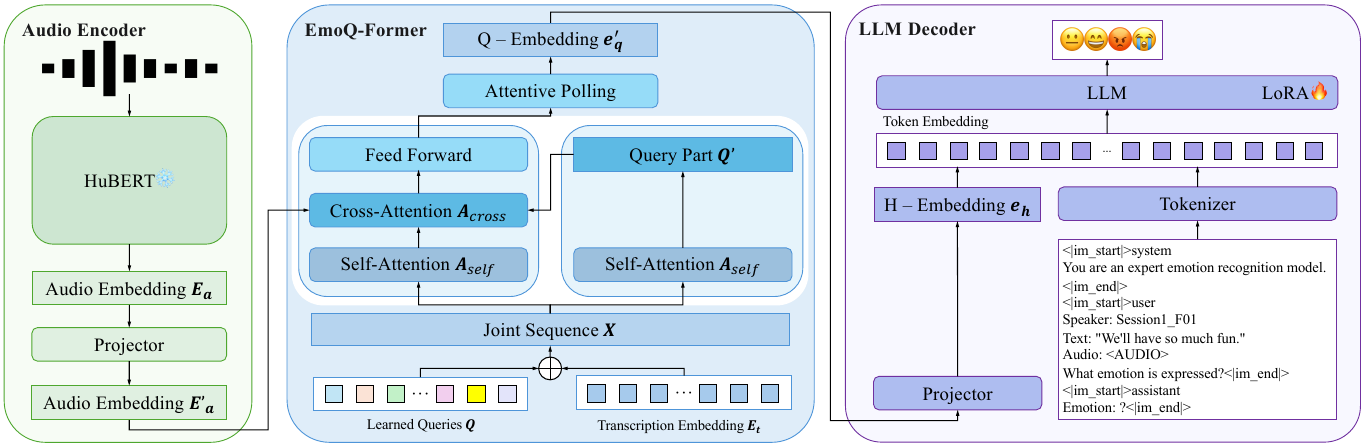}
\caption{The architecture of EmoQ. It adopts end-to-end multimodal learning and contains three core modules: (1) An audio encoder that employs HuBERT to encode the features of speech signals in the form of audio embeddings; (2) an EmoQ-Former that realizes cross-modal fusion of audio and text by learnable queries and an attention masking strategy; and (3) an LLM decoder that uses an LLM to map the fused multimodal representations and tokenized prompts to emotion categories.}
\label{fig:detailed_emoq}
\end{figure*}

EmoQ is based on the MLLM framework as shown in Fig. \ref{fig:detailed_emoq}. The framework consists of an audio encoder, a Q-Former-based bridge-net, and an LLM decoder. The EmoQ-Former is inspired by BLIP-2~\cite{li2023blip2}, and it is responsible for aligning acoustic and textual features to generate a fused representation. This representation is injected into the LLM as soft-prompts to determine the emotion state of an input utterance.

\subsection{EmoQ-Former}
Before entering the EmoQ-Former, we preprocess the input for the audio and text modalities. For the audio modality $a$, we use the pre-trained HuBERT model~\cite{hsu2021hubert} to extract the audio embeddings $\bm{E}_a \in \mathbb{R}^{L_a \times d_a}$, where $L_a$ is the number of audio frames and $d_a$ is the feature dimension. We map the audio embeddings to $\bm{E}'_a \in \mathbb{R}^{L_a \times d_h}$ by a linear projection layer, aligned with the hidden dimension $d_h$ of the query vectors. For the corresponding transcribed text $t$, we use a tokenizer to convert it to a text embedding sequence $\bm{E}_t \in \mathbb{R}^{L_t \times d_h}$, where $L_t$ is the number of tokens.

We concatenate the learned query vectors $\bm{Q}\in \mathbb{R}^{N_q \times d_h}$ with the text embeddings $\bm{E}_t$ to form a joint sequence $\bm{X}$:
\begin{equation} 
\bm{X} = \text{Concat}(\bm{Q}, \bm{E}_t) \in \mathbb{R}^{(N_q + L_t) \times d_h}.
\end{equation}
This sequence is processed by a self-attention mechanism such that the query vectors and text embeddings can attend to each other, allowing the query vectors to absorb the semantic information. Specially, we have:
\begin{equation} 
\bm{A}_{\text{self}} = \text{Softmax}\left(\frac{(\bm{X}\bm{W}_Q)(\bm{X}\bm{W}_K)^\top}{\sqrt{d_k}}\right)(\bm{X}\bm{W}_V),
\end{equation}
where $\bm{W}_Q$, $\bm{W}_K$, and $\bm{W}_V$ are the weight matrices of the self-attention mechanism, and $d_k$ is the dimension of the key vectors. Next, we feed the matrix $\bm{Q}' \in \mathbb{R}^{N_q \times d_h}$ , which is the query portion of the self-attended matrix $\bm{A}_{\text{self}}$, into the cross-attention module, allowing the vectors in this portion to attend to the projected audio embeddings $\bm{E}'_a$:
\begin{align} 
\bm{A}_{\text{cross}} &= \text{Softmax}\left(\frac{(\bm{Q}'\bm{W}_{Q,\text{cross}})(\bm{E}'_a\bm{W}_{K,\text{cross}})^\top}{\sqrt{d_k}} \odot \bm{M}\right) \nonumber \\
&\quad \times (\bm{E}'_a\bm{W}_{V,\text{cross}}),
\end{align}
where $\bm{W}_{Q,\text{cross}}$, $\bm{W}_{K,\text{cross}}$, and $\bm{W}_{V,\text{cross}}$ are the weight matrices of the cross-attention, and $\bm{M} \in \mathbb{R}^{N_q \times L_a}$ is an attention mask matrix to handle variable-length audio sequences. Specifically, $M_{1:N_q,j}=1$ if the time position $j$ corresponds to valid audio content; otherwise, $M_{1:N_q,j}=0$ to accommodate padded time positions beyond the actual audio length. We then present the cross-attended vectors $\{\bm{A}_{\text{cross},i}\}^{N_q}_{i=1} \in \mathbb{R}^{d_h}$ to a feed forward network (FFN), followed by passing the resulting vectors to an attentive pooling layer with multiple attention heads to obtain $\bm{e}_q$. Finally, we apply L2-norm to normalize $\bm{e}_q$ to obtain the final embedding $\bm{e}'_q$.

\subsection{Multi-objective Affective Learning (MAL)}
We propose a multi-objective affective learning framework that combines supervised contrastive loss~\cite{khosla2020supervised} and focal loss~\cite{lin2017focal} to make the cross-modal embeddings more discriminative and to handle the class imbalance issue in speech emotion recognition. Supervised contrastive loss leverages the label information to reinforce the principle that samples of the same class will be close to each other in the representation space and that samples from different classes will be far apart. It enhances the quality of the feature representation by using a variant of InfoNCE loss:
\begin{equation} 
\mathcal{L}_{\textit{SCL}} = -\frac{1}{N}\sum_{i=1}^{N}\frac{1}{|P(i)|}\sum_{j \in P(i)} \log\frac{\exp(\bm{e}_i \cdot \bm{e}_j / \tau)}{\sum_{k \in A(i)}\exp(\bm{e}_i \cdot \bm{e}_k / \tau)},
\end{equation}
where $\bm{e}_i$ denotes the L2-normalized cross-modal feature vector $\bm{e}'_q$ in Fig.~\ref{fig:detailed_emoq} for the $i$-th sample in a mini-batch of size $N$. $P(i)$ contains the indexes of positive samples with the same emotion label as sample $i$, $A(i)$ contains the indexes of all samples in the batch except sample $i$, and $\tau$ is a temperature parameter. 

To solve the category imbalance problem in emotion classification, we use the focal loss to reduce the contribution of easy-to-categorize samples and enhance the learning weights of difficult samples. For each sample $i$ with embedding $\bm{e}_i$  
and true label $y_i \in \{1,...,C\}$ where $C$ is the number of emotion classes, we compute the logits $\bm{z}_i \in \mathbb{R}^C$ through a classification head to obtain the posterior probability vector $\bm{p}_i \in \mathbb{R}^C$. The focal loss is computed as:
\begin{equation} 
\mathcal{L}_\textit{focal} = -\frac{1}{N}\sum_{i=1}^{N} \alpha_{y_i}(1-p_{i,y_i})^{\gamma}\log(p_{i,y_i}),
\label{eq:focal}
\end{equation}
where $p_{i,y_i}$ is the $y_i$-th element of $\bm{p}_i$, $\alpha_{y_i} \in [0,1]$ is the pre-computed weight for class $y_i$, and $\gamma$ is the focus parameter.

MAL achieves the enhancement of emotion information by multi-objective learning:
\begin{equation} 
\mathcal{L}_\textit{MAL}= \mathcal{L}_\textit{SCL}+\lambda \mathcal{L}_\textit{focal}, 
\label{eq:mal}
\end{equation}
where $\lambda$ is a balancing factor.

\subsection{MLLM Instruction Tuning for Emotion Recognition}
For each sample $i$, we map $\bm{e}'_q$ to a representation $\bm{e}_h \in \mathbb{R}^{d_l}$ that is consistent with the dimensionality of the LLM word-embedding space through a projector. We then design a structured prompt template that instructs the LLM to act as an emotion recognition system. The user input section contains the speaker information (if available), the transcribed text, and the multimodal features represented by a special \texttt{<AUDIO>} placeholder. The template also concludes with an assistant response format that begins with “Emotion:” to ensure consistent output generation.  Fig. \ref{fig:detailed_emoq} (right panel) gives an example of the prompt template. Before feeding the text into the LLM, we replace the embedding vector of the \texttt{<AUDIO>} placeholder with the projected vector $\bm{e}_h$. 

We define the fine-tuning process of the LLM as a standard autoregressive language modeling task. The goal of the model is to maximize the probability of generating a sequence of correct responses given the input context:
\begin{equation} 
\mathcal{L}_{\text{CE}} = -\frac{1}{N} \sum_{i=1}^{N} \sum_{j=1}^{L} w_{ij} \log P(t_{i,j} | t_{i,<j}, \bm{x}_i),
\label{eq:ce_loss}
\end{equation}
where $N$ is the batch size and $L$ is the length of the target sequence. $\bm{x}_i$ is the complete input sequence containing soft-prompts, $\bm{e}_h$, and the token embeddings, and $t_{i,j}$ is the true token of sample $i$ at position $j$. $P(t_{i,j} | t_{i,<j}, \bm{x}_i)$ is the conditional probability of generating the correct token. In order to make the model focus on learning emotion keywords, we introduce the weights $w_{ij}$'s at the token level. When $t_{i,j}$ is a predefined emotion word, its weight is set to a high value. During inference, we extract the logits of the predefined emotion words from the hidden state $\bm{h}_{T,i}$ of the last token of sample $i$'s answer sequence. We compute the logits over the entire vocabulary of the LLM:
\begin{equation}
\bm{l}_i = \bm{h}_{T,i} \bm{W}_o \in \mathbb{R}^{|V|},
\end{equation}
where $\bm{W}_o$ is the output vocabulary projection matrix of the LLM, and $|V|$ is the vocabulary size. We then extract the logits corresponding to the $C$ emotion categories:
\begin{equation} 
\bm{q}_{i} = \text{Softmax}([\bm{l}_{i,k_1}, \bm{l}_{i,k_2}, ..., \bm{l}_{i,k_C}]) \in \mathbb{R}^C,
\end{equation}
where $k_c$ denotes the vocabulary index of the $c$-th emotion word, and $\bm{q}_i$ contains posterior probabilities of the $C$ emotion states given the $i$-th utterance.

\subsection{Training Process}
We designed a two-stage training process. First, we focus on the training of EmoQ-Former to equip it with multimodal feature fusion capability. Second, we align the multimodal information with the semantic space of the LLM to fine-tune the whole system in an end-to-end manner.

The goal of the first phase is to train the EmoQ-Former to be able to fuse acoustic features $\bm{E}_a'$ and text embeddings $\bm{E}_t$ into a highly discriminative multimodal embedding $\bm{e}'_q$. At this stage, we freeze the parameters of the pre-trained HuBERT. The output $\bm{e}'_q$ of the EmoQ-Former is fed into a lightweight auxiliary MLP for emotion classification. The EmoQ-Former and the auxiliary MLP head are optimized by our proposed MAL (Eq.~\ref{eq:mal}). After training, we save the weights of EmoQ-Former and discard the auxiliary MLP head.

In the second stage, we load the EmoQ-Former's weights pre-trained in the first stage. Then, the EmoQ-Former is connected to the projector and the LLM to constitute the complete end-to-end model in Fig. \ref{fig:detailed_emoq}. We use Low-Rank Adaptation (LoRA)~\cite{hu2022lora} for fine-tuning of the LLM, where the parameters of the EmoQ-Former and projector are co-trained with the LoRA parameters of the LLM. The whole model (EmoQ-Former and LLM) uses the $\mathcal{L}_{\text{CE}}$ in Eq.~\ref{eq:ce_loss} as the optimization objective.

\section{Experiments}
\label{sec:typestyle}
\subsection{Datasets details}
The IEMOCAP~\cite{busso2008iemocap} dataset is an authoritative benchmark dataset in the field of multimodal affective computing, which consists of conversation clips recorded by 10 native English speakers in a controlled environment. We chose four main emotion categories: neutral, angry, sadness, and happy, covering 5531 utterances. Among them, the category of happy integrates the labels of happy and excited from the original dataset. The dataset is divided using speaker-independent five-fold cross-validation.

The MELD~\cite{poria2019meld} dataset was derived from the dialog clips of the TV series Friends and contains 13,708 utterances. This dataset provides pre-partitioned training, validation, and test sets containing 9989, 1109, and 2610 samples, respectively. We designed two sets of experimental configurations: a four-category experiment with emotion categories neutral, angry, sadness, and joy; and a seven-category experiment that uses the complete emotion labels of the dataset, including neutral, angry, sadness, joy, surprise, disgust, and fear. 
\subsection{Experiment Setup}
The audio encoder used the pre-trained HuBERT-large model, and the parameters were kept frozen during the training process. The EmoQ-Former comes from the pre-trained RoBERTa-base~\cite{liu2019roberta}. The LLM decoder used the Qwen2.5-7B-Instruct~\cite{yang2024qwen2}, and adopted the LoRA fine-tuning strategy, where the rank parameter $r=16$ and dropout rate were set to 0.1. The temperature parameter $\tau$ for $\mathcal{L}_\textit{SCL}$ and the focus parameter for $\mathcal{L}_\textit{focal}$ were set to 0.07 and 2.0, respectively. The class weights $\alpha$'s in Eq.~\ref{eq:focal} were adaptively computed based on the training set's class distribution. The balancing factor $\lambda$ was set to 1 in $\mathcal{L}_\textit{MAL}$. The optimizer was chosen to be AdamW~\cite{loshchilov2019decoupled}, with learning rate set to $1 \times 10^{-5}$ and a batch size of 8. We utilized the officially provided ground-truth transcriptions as the text input. We use a Weighted Accuracy (WA), Unweighted Accuracy (UA), and weighted F1-score as evaluation metrics.

\subsection{Main Results and Analysis}

\begin{table}[t]
\centering
\caption{Performance comparison on IEMOCAP's 4-class emotion recognition task}
\label{tab:iemocap_results}
\begin{tabular}{lccc}
\toprule
\textbf{Model} & \textbf{Year} & \textbf{WA (\%)} & \textbf{UA (\%)} \\
\midrule
HuBERT large~\cite{emobox} & 2021 & 66.7 & 67.4 \\
Co-attention~\cite{coattention} & 2022 & 69.8 & 71.1 \\
SpeechFormer++~\cite{speechformer} & 2023 & 70.5 & 71.5 \\
MSTR~\cite{MSTR} & 2023 & 70.6 & 71.6 \\
ENT~\cite{ENT} & 2024 & 72.4 & 73.0 \\
MFSN~\cite{MFSN} & 2024 & 71.9 & 74.0 \\
Vesper-12~\cite{vesper} & 2024 & 70.7 & 70.8 \\
\midrule
\textbf{EmoQ (Ours)} & \textbf{2025} & \textbf{74.4} & \textbf{74.5} \\
\bottomrule
\end{tabular}
\end{table}

\begin{table}[t]
\centering
\caption{Performance comparison on MELD's 4-class emotion recognition task}
\label{tab:meld_4class_results}
\small
\begin{tabular}{lcccc}
\toprule
\textbf{Model} & \textbf{Year} & \textbf{WA (\%)} & \textbf{UA (\%)} & \textbf{WF1 (\%)} \\
\midrule
HuBERT large~\cite{emobox} & 2021 & 46.4 & 24.1 & - \\
MetricAug~\cite{metric} & 2023 & - & - & 52.9 \\
EAMT~\cite{EAMT} & 2023 & - & 45.2 & 44.4 \\
FMASR~\cite{FMASR} & 2024 & - & 56.3 & - \\
GSER~\cite{GSER} & 2024 & 63.0 & - & - \\
LLM-CL~\cite{LLM-CL} & 2025 & 61.4 & 52.6 & 59.9 \\
\midrule
\textbf{EmoQ (Ours)} & \textbf{2025} & \textbf{73.6} & \textbf{57.2} & \textbf{71.6} \\
\bottomrule
\end{tabular}
\begin{tablenotes}
\small
\item "-" indicates the metric was not reported in the original paper.
\end{tablenotes}
\end{table}
\begin{table}[t]
\centering
\caption{Performance comparison on MELD's 7-class emotion recognition task}
\label{tab:meld_7class_results}
\small
\begin{tabular}{lcccc}
\toprule
\textbf{Model} & \textbf{Year} & \textbf{WA (\%)} & \textbf{UA (\%)} & \textbf{WF1 (\%)} \\
\midrule
MSTR~\cite{MSTR} & 2023 & - & - & 46.2 \\
SpeechFormer++~\cite{speechformer} & 2023 & 51.0 & 27.3 & 47.0 \\
DST~\cite{DST} & 2023 & - & - & 48.8 \\
DWFormer~\cite{dwformer} & 2023 & - & - & 48.5 \\
Vesper-12~\cite{vesper} & 2024 & 53.5 & 26.8 & 48.0 \\
DropFormer~\cite{dropformer} & 2024 & - & - & 49.3 \\
\midrule
\textbf{EmoQ (Ours)} & \textbf{2025} & \textbf{67.6} & \textbf{50.8} & \textbf{66.5} \\
\bottomrule
\end{tabular}
\end{table}

On the IEMOCAP four-category task, as shown in Table \ref{tab:iemocap_results}, EmoQ improves over the best model by 2.0 and 0.5 percentage points on WA and UA respectively. On the more challenging MELD dataset, EmoQ shows even greater advantages. As shown in Table \ref{tab:meld_4class_results}, our model achieves 73.6\% WA and 71.6\% WF1. In the seven-category task (Table \ref{tab:meld_7class_results}), EmoQ leads by 14.1, 23.5, and 17.2 percentage points in WA, UA, and WF1, respectively. 

EmoQ shows significantly greater improvement on MELD than on IEMOCAP, a phenomenon that stems from two designs. In the first stage of training, EmoQ-Former learns a highly discriminative cross-modal emotional representation via the auxiliary classification head, ensuring that the multimodal representation is
suitable for the emotion recognition task and provides emotional cues for the LLM to provide correct responses. In the second stage of training, we provided the LLM with both the complete transcribed text and the emotional soft-prompt representation described above. 

\subsection{Ablation Study}

\begin{table}[t]
\centering
\caption{Effect of input modality and alignment module on MELD's 7-class SER}
\label{tab:architecture_ablation}
\setlength{\tabcolsep}{4pt} 
\begin{tabular}{cc|c|ccc}
\toprule
\multicolumn{2}{c|}{\textbf{Input Modality}} & \multirow{2}{*}{\textbf{EmoQ-Former}} & \multirow{2}{*}{\textbf{WA (\%)}} & \multirow{2}{*}{\textbf{UA (\%)}} & \multirow{2}{*}{\textbf{WF1(\%)}} \\
\cline{1-2}
\textbf{Audio} & \textbf{Text} & & & & \\
\midrule
\checkmark & \ding{55} & \ding{55}  & 47.8 & 26.3 & 43.6 \\
\ding{55} & \checkmark & \ding{55}  & 63.7 & 42.9 & 60.0 \\
\checkmark & \checkmark & \ding{55} & 64.5 & 45.3 & 61.7 \\
\checkmark & \checkmark & \checkmark & \textbf{67.6} & \textbf{50.8} & \textbf{66.5} \\
\bottomrule
\end{tabular}
\begin{tablenotes}
\small
\item \checkmark~indicates the component is used; \ding{55}~indicates the component is not used.
\end{tablenotes}
\end{table}

\begin{table}[t!]
\centering
\caption{Impact of MAL (Eq.~\ref{eq:mal}) on MELD's 7-Class SER}
\label{tab:mal_ablation}
\begin{tabular}{cc|ccc}
\toprule
\textbf{SCL} & \textbf{Focal Loss} & \textbf{WA (\%)} & \textbf{UA (\%)} & \textbf{WF1 (\%)} \\
\midrule
\ding{55} & \ding{55} & 62.3 & 41.2 & 58.2 \\
\ding{55} & \checkmark & 65.7 & 49.4 & 64.2 \\
\checkmark & \ding{55} & 64.0 & 44.6 & 60.8 \\
\checkmark & \checkmark & \textbf{67.6} & \textbf{50.8} & \textbf{66.5} \\
\bottomrule
\end{tabular}
\end{table}

\begin{table}[t!]
\centering
\caption{Benefit of Stage-1 pre-training on MELD's 7-class SER}
\label{tab:training_ablation}
\begin{tabular}{c|ccc}
\toprule
\textbf{Stage-1 Pre-training} & \textbf{WA (\%)} & \textbf{UA (\%)} & \textbf{WF1(\%)} \\
\midrule
\ding{55} & 64.9 & 50.6 & 63.3 \\
\checkmark & \textbf{67.6} & \textbf{50.8} & \textbf{66.5} \\
\bottomrule
\end{tabular}
\end{table}

We verified the effect of different input modalities on the performance of SER, as shown in Table~\ref{tab:architecture_ablation}. Four configurations were set up for the experiments: Audio-only, Text-only, and Audio and Text without and with EmoQ-Former. The Audio-only configuration directly inputted the audio features $\bm{E}_a$ extracted by HuBERT into the LLM decoder and skipped the EmoQ-Former module. Text-only configuration used only the transcribed text $t$ as input and performed emotion categorization by the prompt template, obtaining 63.7\% WA and 42.9\% UA. Audio and Text without EmoQ-Former fed audio features and text features into the LLM via prompt and soft injection, and the performance improves to 64.5\% WA and 45.3\% UA. The complete Audio and Text with EmoQ-Former generated a fusion representation $\bm{e}'_q$ via EmoQ-Former, which was mapped to $\bm{e}_h$ by the projector and injected into LLM, obtaining 67.6\% WA and 50.8\% UA. This demonstrates the effectiveness of EmoQ-Former in cross-modal feature fusion.

We analyzed the contribution of each component in the MAL framework to the model performance, as detailed in Table~\ref{tab:mal_ablation}. Both focal loss and SCL 
individually improve performance over the baseline, with 
focal loss being particularly effective against the class 
imbalance of the MELD dataset. The full MAL 
configuration combining both losses achieves the best 
performance.

The effectiveness of the two-stage training strategy is verified by comparing the end-to-end training against the two-stage training, as shown in Table~\ref{tab:training_ablation}. End-to-end training (1st row) directly optimizes the full model, whereas two-stage training (2nd row) pre-trains the EmoQ-Former via an auxiliary MLP head followed by LLM fine-tuning. Evidently, the results show that two-stage training is superior, as it allows the EmQ-Former to learn a high-quality cross-modal representation for the LLM model to produce correct responses for the input prompts.

\section{Conclusion}
\label{sec:majhead}
In this paper, we propose the EmoQ framework, which effectively fuses audio information and textual semantics through EmoQ-Former and leverages multi-objective affective learning. Experimental results show that EmoQ achieves state-of-the-art performance on both IEMOCAP and MELD datasets. This study demonstrates the potential of MLLMs in emotion recognition tasks.


\begin{thebibliography}{31}
\bibitem{hsu2021hubert}
W.-N. Hsu, B. Bolte, Y.-H. H. Tsai, K. Lakhotia, R. Salakhutdinov, and A. Mohamed,
``HuBERT: Self-supervised speech representation learning by masked prediction of hidden units,''
\emph{IEEE/ACM Trans. Audio, Speech, and Language Processing}, vol. 29, pp. 3451--3460, 2021.
\bibitem{li2023blip2}
J. Li, D. Li, S. Savarese, and S. Hoi,
``BLIP-2: Bootstrapping language-image pre-training with frozen image encoders and large language models,''
in \emph{Proc. Int. Conf. Mach. Learn. (ICML)}, 2023.
\bibitem{hu2022lora}
E. J. Hu, Y. Shen, P. Wallis, Z. Allen-Zhu, Y. Li, S. Wang, L. Wang, and W. Chen,
``LoRA: Low-rank adaptation of large language models,''
in \emph{Proc. Int. Conf. Learn. Represent. (ICLR)}, 2022.
\bibitem{lin2017focal}
T.-Y. Lin, P. Goyal, R. Girshick, K. He, and P. Doll{\'a}r,
``Focal loss for dense object detection,''
\emph{IEEE Trans. Pattern Anal. Mach. Intell. (TPAMI)}, vol. 42, no. 2, pp. 318--327, 2020.
\bibitem{liu2019roberta}
Y. Liu, M. Ott, N. Goyal, J. Du, M. Joshi, D. Chen, O. Levy, M. Lewis, L. Zettlemoyer, and V. Stoyanov,
``RoBERTa: A robustly optimized BERT pretraining approach,''
\emph{arXiv preprint arXiv:1907.11692}, 2019.
\bibitem{yang2024qwen2}
Qwen et al.,
``Qwen2.5 technical report,''
\emph{arXiv preprint arXiv:2412.15115}, 2025.
\bibitem{busso2008iemocap}
C. Busso, M. Bulut, C.-C. Lee, A. Kazemzadeh, E. Mower, S. Kim, J. N. Chang, S. Lee, and S. S. Narayanan,
``IEMOCAP: Interactive emotional dyadic motion capture database,''
\emph{Language Resources and Evaluation}, vol. 42, no. 4, pp. 335--359, 2008.
\bibitem{poria2019meld}
S. Poria, D. Hazarika, N. Majumder, G. Naik, E. Cambria, and R. Mihalcea,
``MELD: A multimodal multi-party dataset for emotion recognition in conversations,''
in \emph{Proc. 57th Annual Meeting of the Association for Computational Linguistics (ACL)}, 2019, pp. 527--536.
\bibitem{khosla2020supervised}
P. Khosla, P. Teterwak, C. Wang, A. Sarna, Y. Tian, P. Isola, A. Maschinot, C. Liu, and D. Krishnan,
``Supervised contrastive learning,''
in \emph{Advances in Neural Information Processing Systems (NeurIPS)}, 2020, pp. 18661--18673.
\bibitem{loshchilov2019decoupled}
I. Loshchilov and F. Hutter,
``Decoupled weight decay regularization,''
in \emph{Proc. Int. Conf. Learn. Represent. (ICLR)}, 2019.
\bibitem{mmnodeformer}
Z. Huang, M.-W. Mak, and K. A. Lee,
``MM-NodeFormer: Node Transformer multimodal fusion for emotion recognition in conversation,''
in \emph{Proc. Interspeech}, 2024, pp. 4069-4073.
\bibitem{emobox}
Z. Ma, M. Chen, H. Zhang, Z. Zheng, W. Chen, X. Li, J. Ye, X. Chen, and T. Hain,
``EmoBox: Multilingual multi-corpus speech emotion recognition toolkit and benchmark,''
in \emph{Proc. Interspeech}, 2024.
\bibitem{coattention}
H. Zou, Y. Si, C. Chen, D. Rajan, and E. S. Chng,
``Speech emotion recognition with co-attention based multi-level acoustic information,''
in \emph{Proc. IEEE Int. Conf. Acoust., Speech and Signal Process. (ICASSP)}, 2022, pp. 7367--7371.
\bibitem{speechformer}
W. Chen, X. Xing, X. Xu, J. Pang, and L. Du,
``SpeechFormer++: A hierarchical efficient framework for paralinguistic speech processing,''
\emph{IEEE/ACM Trans. Audio, Speech, and Language Processing}, vol. 31, pp. 775--788, 2023.
\bibitem{MSTR}
Z. Li, X. Xing, Y. Fang, W. Zhang, H. Fan, and X. Xu,
``Multi-scale temporal transformer for speech emotion recognition,''
in \emph{Proc. Interspeech}, 2023, pp. 3652--3656.
\bibitem{ENT}
S. Shen, Y. Gao, F. Liu, H. Wang, and A. Zhou,
``Emotion neural transducer for fine-grained speech emotion recognition,''
in \emph{Proc. IEEE Int. Conf. Acoust., Speech and Signal Process. (ICASSP)}, 2024, pp. 10111--10115.
\bibitem{MFSN}
H. Sun, F. Zhang, Y. Gao, S. Zhang, Z. Lian, and J. Feng,
``MFSN: Multi-perspective fusion search network for pre-training knowledge in speech emotion recognition,''
in \emph{Proc. Interspeech}, 2024, pp. 4703--4707.
\bibitem{metric}
Y.-T. Wu and C.-C. Lee,
``MetricAug: A distortion metric-lead augmentation strategy for training noise-robust speech emotion recognizer,''
in \emph{Proc. Interspeech}, 2023, pp. 3587--3591.
\bibitem{EAMT}
X. Shi, X. Li, and T. Toda,
``Emotion awareness in multi-utterance turn for improving emotion prediction in multi-speaker conversation,''
in \emph{Proc. Interspeech}, 2023, pp. 765--769.
\bibitem{FMASR}
T. Feng and S. Narayanan,
``Foundation model assisted automatic speech emotion recognition: Transcribing, annotating, and augmenting,''
in \emph{Proc. IEEE Int. Conf. Acoust., Speech and Signal Process. (ICASSP)}, 2024, pp. 12116--12120.
\bibitem{GSER}
A. Ibrahim, S. Shehata, A. Kulkarni, M. Mohamed, and M. Abdul-Mageed,
``What does it take to generalize SER model across datasets? A comprehensive benchmark,''
in \emph{Proc. Interspeech}, 2024, pp. 1590--1594.
\bibitem{LLM-CL}
H. Zou, F. Lv, D. Zheng, E. S. Chng, and D. Rajan,
``Large language models meet contrastive learning: Zero-shot emotion recognition across languages,''
\emph{arXiv preprint arXiv:2503.21806}, 2025.
\bibitem{vesper}
W. Chen, X. Xing, P. Chen, and X. Xu,
``Vesper: A compact and effective pretrained model for speech emotion recognition,''
\emph{IEEE Transactions on Affective Computing}, vol. 15, no. 3, pp. 1711--1724, 2024.
\bibitem{DST}
W. Chen, X. Xing, X. Xu, J. Pang, and L. Du,
``DST: Deformable speech transformer for emotion recognition,''
in \emph{Proc. IEEE Int. Conf. Acoust., Speech and Signal Process. (ICASSP)}, 2023, pp. 1--5.
\bibitem{dwformer}
S. Chen, X. Xing, W. Zhang, W. Chen, and X. Xu,
``DWFormer: Dynamic window transformer for speech emotion recognition,''
in \emph{Proc. IEEE Int. Conf. Acoust., Speech and Signal Process. (ICASSP)}, 2023, pp. 1--5.
\bibitem{dropformer}
J. Mai, X. Xing, W. Chen, and X. Xu,
``DropFormer: A dynamic noise-dropping transformer for speech emotion recognition,''
in \emph{Proc. Interspeech}, 2024, pp. 2645--2649.
\bibitem{salmonn}
C. Tang, W. Yu, G. Sun, X. Chen, T. Tan, W. Li, L. Lu, Z. Ma, and C. Zhang,
``SALMONN: Towards generic hearing abilities for large language models,''
in \emph{Proc. Int. Conf. Learn. Represent. (ICLR)}, 2024.
\bibitem{secap}
Y. Xu, H. Chen, J. Yu, Q. Huang, Z. Wu, S.-X. Zhang, G. Li, Y. Luo, and R. Gu,
``SECap: Speech emotion captioning with large language model,''
in \emph{Proc. AAAI Conf. on Artificial Intelligence}, vol. 38, no. 17, pp. 19323--19331, 2024.
\bibitem{zhu2024dca}
Q. Zhu, C. Zheng, Z. Zhang, W. Shao, and D. Zhang,
``Dynamic Confidence-Aware Multi-Modal Emotion Recognition,''
\emph{IEEE Transactions on Affective Computing}, vol. 15, no. 3, pp. 1358--1370, 2024.
\bibitem{yin2024survey}
S. Yin, C. Fu, S. Zhao, K. Li, X. Sun, T. Xu, and E. Chen,
``A Survey on Multimodal Large Language Models,''
\emph{arXiv preprint arXiv:2306.13549}, version 4, Nov. 2024.
\end{thebibliography}
\end{document}